\documentclass{ws-jaic}
\usepackage{ws-jaic-thm}        
\usepackage[square]{natbib}
\usepackage{graphicx}
\usepackage{float}

\newcommand\ditem[1]{\item{\bfseries #1}\\}
\begin{document}

\markboth{Mark Hadley}{A generic model of consciousness}


\title{A generic model of consciousness}

\author{Mark Hadley\footnote{DrMarkHadley@gmail.com}}




\maketitle


\begin{abstract}
This is a model of consciousness. The hard problem of consciousness, what it feels like, is answered. The work builds on medical research analyzing the source and mechanisms associated with our feelings. It goes further by describing a generic model with wide applicability. The model is fully consistent with medical pathways in humans, but easily extends to animals and AI. The essence of the model is the interplay between associative memory and physiology. The model is a clear and concrete counterexample to the famous philosophical objections to a scientific explanation. 

\end{abstract}

\section{Introduction}	
We are going to construct a model of “what it feels like”. The work is motivated by the philosophical debates about consciousness. A dominant theme in the philosophy literature is that the intractable problem of consciousness is to explain what it feels like. Equivalent terms are “what it feels like”, “subjective experience”, “conscious experience”, “consciousness”, “qualia”, “phenomenal experience”. This work presents a generic model of what it feels like, informed by medical science, but with much wider applicability. The model can assess artificial consciousness and and design consciousness into AI.

The value of having a generic model cannot be overstated. Current questions about AI and consciousness cannot be answered or informed by anatomical studies of human beings alone. The pathways and components will be different, non-biological. This limitation is even true for conclusions about other animals. If an organ or piece of anatomy is different to a human one, how can we decide if it is relevant or materially significant? This generic model is the answer, it describes the functionality to implement “what it feels like” and is independent of the platform, animal, or agent. 

Consciousness is a challenging subject, not least because there are several quite different uses of the word consciousness, and the term is frequently invoked without definition. Consciousness is an interdisciplinary topic and a major research front in psychology and neuroscience as well as philosophy. \citet{chalmers1995} is an influential introduction to the philosophical side of the subject. He draws a clear distinction between the so-called easy problems which science knows how to approach and is indeed making progress on: 
\begin{quote}
 “The easy problems of consciousness include those of explaining the following phenomena: the ability to discriminate, categorize, and react to environmental stimuli; the integration of information by a cognitive system; the reportability of mental states; the ability of a system to access its own internal states; the focus of attention; the deliberate control of behavior; the difference between wakefulness and sleep …… All of them are straightforwardly vulnerable to explanation in terms of computational or neural mechanisms. ….
…. The really hard problem of consciousness is the problem of experience”.
\end{quote}

\citet{chalmers1997}[p4]  introduces subjective experience:
\begin{quote}
“When we perceive, think and act, there is a whir of causation and information processing …… There is also an internal aspect; there is something it feels like to be a cognitive agent. This internal aspect is conscious experience.”
\end{quote}

We are tackling the hard problem but are not endorsing it. Other commentators do not accept Chalmers’ categorization and have conducted research to show that the so-called hard problem is amenable to scientific understanding \citep{damasio2013}. Indeed, the model presented here directly contradicts the existence of an intractable hard problem. Despite contrary views and recent research, the notion of a hard problem persists in current debates, so this work adds to the weight of evidence for a scientific explanation of consciousness. But it goes further, by extending the research from human anatomy to a generic model it has wider applicability, answers more questions, and offers greater insight.
 
The equivalent terms used by philosophers for the hard problem are consciousness, experiences, qualia, phenomenology, phenomenal, subjective experience, conscious experience, what it is like, what it feels like. They are used interchangeably. Of interest are the inner feelings that accompany an event - not the information transfers or firing of neurons, but why there is a feeling associated with it; a feeling that we experience in the first person. \citet{nagel1974} paper “What is it like to be a bat?” was most influential. Of course, he is not asking what it is like to be 10cm tall and weigh 150g and hang upside down, he is exploring what it feels like to be a bat. And how impossible it is for humans to imagine or explain. He equates the hard problem of consciousness with understanding the origin of what it feels like to be a bat.
We will model what an experience feels like. There are several reasons to pick the term “what it feels like” as our explanatory challenge:
\begin{romanlist}
\item 
The hard problem of consciousness has been described with the phrase “what it feels like” “or what it is like” consistently, continually, and repeatedly from \cite{nagel1974} through to \citet{chalmers1997} and to the present day. That is consistent for 40 years.
\item
The phrase has not been reinterpreted. No philosophical version of it has been defined or described. It is used as is because we know what it means. There has been no attempt to change the popular meaning. This contrasts with expressions like subjective experience and phenomenal which have been reinterpreted by philosophers specifically for the academic discipline and need to be explained in books for the general public.
\item
It is used in introductory academic texts and in books aimed at the general public. It is the language chosen to relate complex philosophical concepts to the public using language they understand. 
\end{romanlist}

We will model what it feels like and how those feelings arise in response to different stimuli. The model is generic in terms of logical building blocks that can be applied widely. This complements the medical research into the pathways and mechanisms (predominantly in humans). The two approaches are fully compatible, and we will repeatedly cross reference this generic model against medical detail.

\section{Understanding what it feels like}
We are in a good position to understand what it feels like. Contrary to assertions that it encapsulates a hard intractable problem of consciousness, we can discuss, analyze, and investigate it. We can attempt to explain what it feels like because the expression and the response are intimately linked to medical diagnosis. As part of a medical examination doctors will ask, how do you feel, and the patient will respond in a way that gives the doctor information about the patient’s body. Indeed, for some conditions such as fibromyalgia, what it feels like is the dominant contribution to a diagnosis.

In general, we can recognize the connection between what it feels like and the state of our bodies. This can be discussed and analyzed widely and across disciplines. In specific terms medical research can, and does, investigate the pathways that connect our physiology to a sense of what it feels like. These medical pathways, that give us a sense of what it feels like, are being investigated with considerable success. See for example \cite{damasio2013}. Some, but not all, aspects of our physiology are sensed, and signals are returned to central nervous system CNS. The exact parts of the CNS that generate emotions is still an active subject of research but it is clear that more than one area is relevant.

We understand what it means to feel cold. If the individual is in an environment with a low temperature for a prolonged period, the physiology will be a cold one. We have a name for it in the English language. Core body temperature drops. For a human, surface arteries will close. The skin color will change to blue. The skin will have goose pimples (piloerection), and muscles contract to cause shivering. There are other physiological changes known and probably more unknown. We say the individual is cold. Being cold is a language label for that set of physiological conditions. We will use the cold state as an example in the following discussions because being cold is widely experienced, supported by shared language, and has known biological mechanisms. The subset of physiological parameters that equate to feeling like being cold are those which can send stimuli to the brain. The pathways are well established by medical science: temperature is detected by sensory receptors in the skin. This sends a signal via an afferent sensory neurone to the central nervous system and then to the thalamus (part of the brain) via the spinothalamic tract. See for example \cite{thermoreceptors}. Similarly, we know what it means to feel happy, frightened, or hungry. There are many more feelings with distinct mechanisms. Some are well understood, others less so. In detail, the receptors and pathways are varied. The work of Damasio and others gives increasing understanding about why there is a sense of what it feels like.

We take the essential features and encapsulate them in a simple generic model. The first is that we have a body, a physiology. (Damasio and Carvalho emphasize that the body is in homeostasis, this seems not to be fundamental to us, so we use the term physiology.) An enormous number of parameters characterize our physiology. A subset of them have receptors that feed signals back to the neuraxis. It is this subset that gives rise to feelings. When we report against those signals, we report what it feels like. Why we have evolved awareness of some parameters and not others may well have evolutionary origins. And the degree to which other creatures have signals from their physiology to mind, limits the scope of the model and the attribution of conscious feelings. 

This preliminary model is more than just information processing. The physiology is essential. That is a conclusion shared by medical research studying feelings. Already this feature distinguishes the model form speculations about AI and information processing. As Chalmers and others claim for the hard problem of consciousness: a subjective feeling is more than just knowledge. It cannot be conveyed by a transfer of information alone.

The philosophy literature focuses particularly on the feelings associated with sensory perceptions. What does it feel like to see red is a common question claimed to encapsulate the puzzle of human consciousness. Our answer to that is based on associative memory.

\section{Response to Stimuli}
So far, we have been describing how we have a sense of what our bodies feel like. We need a bit more than that. Much of the philosophical debate about subjective experience relates to feelings associated with stimuli. What does it feel like to see red, or to hear a bell ring; these are fundamental to the philosophical concept of a hard problem of consciousness. Indeed, they could be characterized as definitive of the hard problem, (to the extent that a hard problem is defined). To progress, we need to consider how our body responds to stimuli and the key role played by associative memory.

Human memory is associative, in contrast to computer memory which is addressable. Although one type of memory can mimic the other, they are quite different in principle. If we see a tomato, then we recall things associated with a tomato: perhaps the sound of the word, or its spelling. Maybe a recipe etc. The associations and recall are automatic and immediate, they do not arise at the end of a process, unlike a rational or logical analysis.  The associations will have been formed through our lives and are not generally identical. They are subjective. The pattern of our neurons changes to embody the associations; it is a feature of our brains called neuroplasticity. Again, it is in marked contrast to conventional computers which have fixed structures and connections.  For humans, the relationships are learned, they can be taught. Associative memory is powerful in some applications and is being harnessed with artificial neural network computing.

The tomato example above was about information processing. However associative memory does more than just recall information, it can recall physiological states. It is a primary mechanism in animals to respond to a stimulus by getting their body ready for whatever is likely to happen next. This associative memory and the responses are personal in two respects: it affects the person’s physiology, and it can vary from person to person depending on what they have learned from past experiences.

One famous example is Pavlov's dog. Dr Pavlov rang a bell when his dog was fed. After a while the dog would salivate when the bell rang. We do not know what it feels like to be a dog, but we do have some clues. Eating food is pleasurable, and necessary for survival. There are several scientific techniques to measure pleasure in animals such has the level of the dopamine hormone. We know that salivation is a physiological response when eating and we know salivation is nerve mediated. Scientifically we know and understand the mechanisms: the salivary glands are mainly supplied with parasympathetic innervation. The reflex: Afferent impulses are sent to the salivary nuclei in the brain-stem, which in turn then send a signal via the glossopharyngeal nerve to the salivary glands to increase saliva production. See for example:\citet{salivary}. So, we know that some of the physiology of eating is recalled when the dog hears the bell ring. We cannot empathize fully with the dog’s subjective experience because our physiology is different to that of a dog. However, we can approach an understanding of what it feels like, because we have some physiology in common.

Another dog could be beaten whenever the bell rings. The physiological response to the violence would be distinctive and pronounced: adrenaline would be released from the adrenal glands along with cortisol (increases blood pressure and blood sugar and some other stuff). Heart rate and respiratory rate increase, blood is directed towards the muscles to prepare the body for ‘fight or flight’. Fatty (glycogen) stores are broken down to release more glucose into the blood. The very different stimuli that this dog experiences will result in a very different response to the sound of the bell. It will not just think differently. Quite apart from logical rational responses, the physiology will change: the dog will feel frightened. The amygdala part of the brain is known to be involved in the feelings of emotions, markedly fear \citep{amygdala}.

What it feels like when the respective dogs hear the same bell sound is very different. It is a subjective experience, but we can explain it with medical knowledge. By extracting the logical elements, we can create a model of what it feels like. Stimuli excite receptors which in turn trigger associative memory that directly causes changes in physiology.

\section{The Generic Model}
The examples are well known and familiar to us. The medical pathways in humans, and animals, have been studied and identified to a greater or lesser extent. We propose a generic model built with abstract functional units but based on current knowledge and research.

The key feature is that the subject has a body, a physiology. Some signals from the body give information about the state of the body. This set of information is what it feels like. External sensory inputs trigger associative memory that partially recalls a mixture of past physiological states. The state of the body changes directly, and automatically, as a consequence of the structure of the associative memory. It is not an analytical or rational process. The subject senses the change to their body giving meaning to what it feels like.

\section{The model with Subtle Stimuli}
The examples above are gross. Very crude responses to strong stimuli. We stated with these examples because they are common to us all and relatively easy to describe with the English language. But the model applies to, and explains, all subjective experiences. What does it feel like to see red? It is subjective, it is different for different individuals. Even when individuals share the same factual information about red, how they feel can be different - their physiology changes in an individual way. Here it is unlikely to be a dramatic change to fundamental processes but a more nuanced subtle effect. Our associative memory recalls experiences associated with the red stimulus at the eyes. It is a parallel process, many historic experiences will be recalled, and with each one, a subset of physiological parameters will be recalled and moderated. The exact mix will be different for different individuals. The recalled physiology is what red feels like to that individual.

We can readily construct examples of different red feelings. If red is associated with the color of your happy family home or the color of hills near where you were brought up. Then the hormones associated with pleasure and comfort will increase - dopamine, serotonin, and endorphins. Conversely you will be calm because hormones associated with stress are suppressed (cortisol, adrenaline). You will feel relaxed and happy. In contrast you may associate red with danger, blood, and distressful incidents.

As a further level of complexity, the red stimulus is unlikely to be isolated, it will be presented as just one of a mix of stimuli interacting with all your senses. The associative memory will recall a whole cocktail of physiological symptoms, some reinforcing, others contradictory. In medical and psychology research the complexity is unhelpful and experimental design tries to reduce it, but to understand the subtlety of what it feels like the richness needs to be acknowledged.

As another example, consider looking at a picture of a loved one vs a stranger. We all know that distinction and can measure it objectively with hormones concentrations, blood pressure etc. This pathway from stimulus to physiological state is medically well known and testable. It is an example that is clearly subjective and can also vary with time for the same individual.

In summary, what it feels like to see red is a subtle change in physiological parameters such as hormone concentrations. That change is created automatically from the complex patterns of interconnection in the individual’s associative memory.

\begin{figure}[h]
\centering
\includegraphics[scale = 0.5]{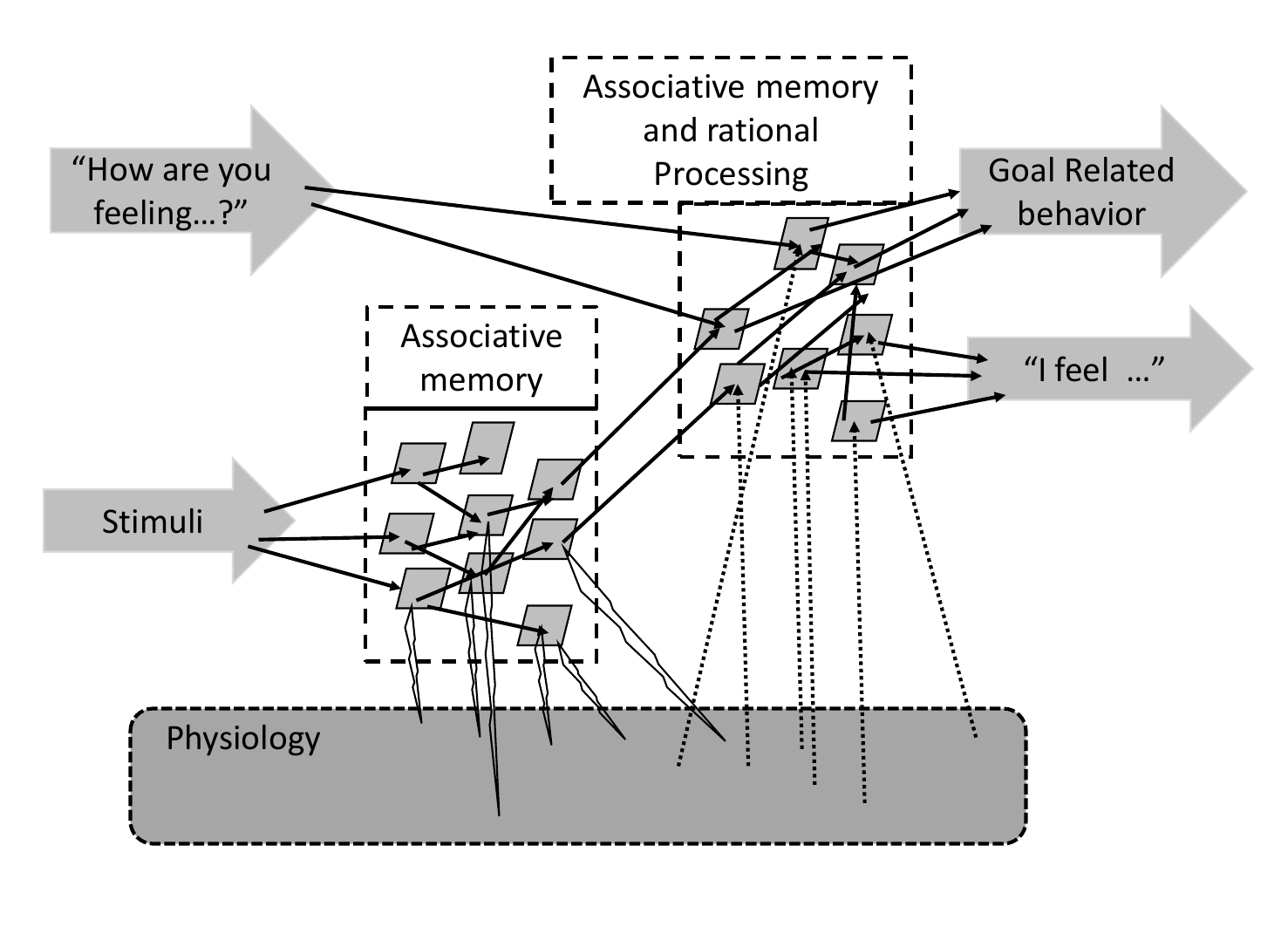}
\caption{ Schematic model of consciousness.\\ Stimuli acting through associative memory, directly influence physiology.\\ Selected signals from the physiology give feelings which can in principle be articulated in a rational way\\ Consciousness will be recognised by behavior and and language influenced by the feelings.}
\end{figure}

\section{Discussion}
We have constructed and described a generic model of what it feels like. We have an implementation of subjective experience which many philosophers consider to be the essence of consciousness. We can now make several illuminating observations

\subsection{Consciousness is graduated}
Consciousness is graded. It depends upon the size of the associative memory, the richness of the physiology and the measure of the interactions between them. The components of the model have a breadth that can vary between agents, creatures, organisms etc. The components can vary in complexity, from humans down to a point where it not so much vanishes but instead has little or no value or interest. This addresses a major deficiency in the philosophy literature where there is no agreement about scope of consciousness.

\subsection{Evolution}
It is sometimes argued that consciousness serves no purpose, we could exist without it. Such an argument adds to the mystery of the subject. We disagree. We have evolved from much simpler mammals with limited ability for rational thought or analysis and certainly unable to handle abstract concepts. But a mouse, for example, must respond to its surroundings. There are times and places where it is appropriate to gather food, eat, sleep, reproduce, fight or flee from danger. The mouse must recognize these places, the environmental signals must be noticed and interpreted. But simple recognition is inadequate, knowing it is in danger is not sufficient for survival. It needs more than information. To survive its body must be ready for the situation. The signals need to trigger appropriate physiological changes. There will be a feeling associated with places and other stimuli. As those feelings and experiences accumulate, they will fashion and adjust links in associative memory to make the mouse more successful.

Conscious experiences, what it feels like, has evolved to allow more primitive creatures to recognise situations and prepare their bodies appropriately. It is essential for survival. It is an implementation of a vital control system that is adaptive and immediately responsive, but does not require rational analysis. It has both preset responses from birth, but associative memory allows the responses to be refined and more effective based on patterns of experience.

\subsection{Goal related feelings}
Humans and animals, presumably through evolution, have primal feelings that are survival related. Consider the obvious: hunger, thirst, fear, pain, arousal. Genetically we try to satisfy hunger and thirst, respond to arousal and minimise fear and pain. The feelings control our behavior. It is the understandable rational behavioral responses in mammals that leads to a belief that they also experience feelings \cite{damasio2013}. 
Nothing in the philosophical description of subjective experience considers subsequent behavior to be definitive or even relevant. For that reason it is not part of the generic model presented here. However, without a doubt any model will be judged by behavioral responses to stimuli. Language descriptions are one part of that, but other aspects of activity are also important. 

\section {AI implementation of consciousness}

Modern computers cannot be conscious in the sense of having conscious experience. It is not a programming issue, it is hardware and operating system dependent. The physiology of a computer is vastly simpler than even the most primitive of insects, and there is no link of associative memory that causes stimuli to recall and reproduce physiological states. That is a hardware requirement. Arguably, some complicated distributed systems could have a significantly rich physiology. Perhaps an airbus 350. but its mechanisms that manipulate the physiology are the result of rational decision making. It does not seem like an appropriate platform to experiment with! But since our model is generic it can apply to even a simple computer system. We will do that to demonstrate the model and expose the challenges of a realistic implementation. A hardware model with a unique operating system is also a good starting point for a simulation.

\subsection{Stimuli}
The stimuli are all the various inputs that the computer responds to. Some interesting ones are listed below:
\begin{itemize}
\item Keyboard strokes
\item Camera and Facial recognition etc
\item  Microphone
\item Time of day  or Ambient light
\end{itemize}
In many respects the set of computer stimuli is as substantial as for humans. Touch, taste and smell are largely missing, but if you add internet connectivity, the range is large, varied and subtle. The missing senses would not be a barrier to consciousness. There is no new hardware required to provide suitable stimuli, normal computer inputs are sufficient and overlap sufficiently with humans to have credibility.

\subsection{The physiology}
For a PC we can identify a few attributes that correspond to physiology. Each in turn can be parameterised in a fairly obvious way and we can draw some playful analogies with human states.
\begin{enumerate}
\ditem{CPU Speed} 
This has some comparisons with heart rate, but also wakefulness.
\ditem{Fan Speed}
As a cooling function it would compare with blood capillary dilation
\ditem{Background processes}
Largely hidden from users a CPU will be running a wide variety of housekeeping type functions. Amongst other things they provide checks and balances and assure smooth running of user initiated processes. There is a loose comparison with bodily processes that maintain homeostasis in mammals.
\ditem{Memory usage}
\ditem{CPU utilisation}
Together with memory usage they indicate a capacity to respond to new challenges. Perhaps related to how much of our bodily functions are taking place and using resources like energy and blood flow.
\ditem{Screen brightness}
The screen is a vital organ for functionality but also a big consumer of energy.
\ditem{Battery level}
The obvious analogy is with energy reserves as fat, but on a shorter timescale would correspond to the energy available to our muscles.
\end{enumerate}
The list is limited and clearly not human nor even mammalian. It is a barrier to the model being accepted. However the limitations are a matter of scale rather than a fundamental limitation. The model will inevitably have more credibility if it is part of a humanoid robot. The physiology would have more familiar aspects and some of the subsequent behavior will seem familiar. 

\subsection{Feelings}
The essence of subjective experience is that the entity has feelings. For humans this is being investigated by neuroscientists; they trace nerve signals that transmit information about some aspects of the physiology to the accessible [rational] part of the brain. The information is a major subset of the physiological parameters listed above but not necessarily all of it.

The model, or demonstration, of artificial consciousness will not work without language, without some articulation of feelings. Not because the language is an essential feature in principle, but because it is the only way to reveal the presence of feelings. The hurdle is high. Talking is a type of behavior and authors such as Chalmers deny that consciousness is defined or characterised by behavior. From the zombie twin argument given below it is clear that some philosophers will never accept a Turing test for consciousness. Although, in a contrary argument, mammals are assumed to be conscious because they behave like us. Neurological studies also reveal sufficient similarities to justify assignments of consciousness to mammals. 

For computers, there is neither a neurological or behavioral similarity to support a claim of consciousness. Quite the contrary, the physical dissimilarities will discourage any such claim. Language is important. Accessible physiological information needs to be named and described. In humans that language is learned and held in associative memory linked to the physiological information. It requires a sense of self-awareness, though perhaps little more than a computer giving system information.  The feelings need to be stated, described and related to functionality. We will give some examples.

How are computer feelings named? We suggest three options, perhaps depending upon the audience. 
\begin{enumerate}
[align = left]
\ditem{Human names}
An obvious choice is to map computer physiology to well known human feelings. Tired,sleepy, excited, headache, alert, nervous ... etc. This will inevitably have limitations in scope and lack precise mapping, but has obvious meaning as a feeling..
\ditem{Computer Jargon}
One could use technical computer expressions and present them as feelings. \emph{Defragmented, high power consumption, high CPU utilisation} etc. These will be manifestly correct, but not necessarily recognised by the public as genuine feelings. Linguistically there is some overlap with human emotions as computer science adopted words with similar meanings. Sleep mode for example.
\ditem{Original words}
Since artificial physiology is distinctly non human, it is appropriate to use different language, made-up words. For example \emph{refranched} when it has recently been scanned and virus checked. Initially this may be an obstacle to humans understanding. However the incomprehensibility of consciousness was famously described by Nagel when he said that the feelings of a bat could never be understood. Using novel language manifestly highlights the first person nature of subjective experience and the differences with human feelings. It may even be critical in winning a philosophical debate.
\end{enumerate}

Human feelings drive behavior and fundamental goals. For a computer the correspondence is limited. Like humans, survival as a functioning entity is critical and should be emphasised. For obvious reasons, the other goal is to provide computing power (in all its dimensions) to users as and when they need it. 

\subsection{Associative link}
Fundamental to the model is that patterns of stimuli and physiological states are recorded in associative memory. This memory is not used to provide information, but to directly recall similar states. Physiology is recalled to varying degrees depending on how well the current stimuli match previous events. It is not a rational, logical or analytical process and it is not algorithmic. This fundamental objection is to some extent undermined because, as has been explained, conventional building blocks can be used to create associative memory.

The operation of the associative link is automatic, more like an operating system interrupt than a normal program. Ctrl-Alt-Del could be compared with a stab of pain creating automatic physiological responses. This directness of operation is important and should be considered as essential for a realistic model. Furthermore, an important feature of the associative link is that it learns. The learning mechanism serves to give more subtle feelings to a wider range of stimuli. It has an increasing memory of stimuli and the related, usually subsequent, physiology. All patterns of stimuli are pattern matched to the current stimuli to influence the physiology. The influence will vary in type and magnitude depending on the degree the patterns match. In a sense we are describing a computer where the operating system itself monitors all incoming stimuli and learns.

We can give some illustrative examples. A user frequently logs on to the computer at about 10pm and starts an intensive spell of gaming. It is CPU intensive, high processor speeds are required and memory demands are high. The computer warms up and cooling fan speeds are increased. You could give that set of parameters a name. Associated linguistic terms could be: In human terms \emph{excited} computer terms \emph{ preparing for high utilisation} and perhaps a new adjective \emph{narksious}. Now that a pattern has been established, facial recognition and the time of day would trigger pre-emptive changes: clearing out unnecessary functions, raising clock speed and turning on cooling fans. There is now a feeling related to the sight of the user at that time of day. An analogy might be adrenaline before participating in a sports match.  As you can see the response has some value and aspects might already be programmed in a rational algorithmic way. However, with this construction any response is learnt and becomes automatic.

Maybe a user regularly takes the computer with them on the train. CPU activity is low but the work continues for an extended period until low battery forces powersaving modes and then a shutdown. Once the pattern is established, the associative memory would link user. location and perhaps time of day and invoke some powersaving modes before they are needed. In language terms it has similarities to adrenalin flow and \emph{anxiety}, in computer terms it is obviously \emph{power saving} or for a new term perhaps \emph{strential}.

The first example is linked to the goal of providing computing power to the user. The second is also a survival function. Other computer goals reside in user programs as part of the rational processing. To extend the model these goals would need to correspond with a physiological parameter described by a feeling. This is either a major new hardware requirement or alternatively requires a dedicated system function to monitor the goal. The system function then needs to be presented a part of the physiology. The latter would inevitably blur the line between user algorithms and feelings.

\section{AI simulation of consciousness}
Stepping back from an implementation, simulating consciousness should be straightforward. Certainly, a substantial amount of work, but the components and structure should now be clear. It should be possible to produce a simulation indistinguishable from a real consciousness.

An important choice is between simulating human consciousness, machine hardware consciousness, animal consciousness or a deliberately alien consciousness.  In each case the starting point is to define a full set of physiology components with parametrised activity levels. Secondly, define a set of feelings and language terms related to the physiological states.

The richness, and one subjective aspect, comes from the associations between stimuli and the recall of past states. Ideally this should be a learned association starting from initial primitive survival related responses. It is unlikely that these associations can be harvested from the internet because that lacks the individuality. Stimuli invoke different feelings in different people. An average over the whole population is unlikely to be effective.

The simulation can then receive a stimulus, recall a past physiological state in whole or part, and describe the state. As a simulation it could also independently report on the physiology. Humans know when they are happy, but cannot give a numerical value to their dopamine concentration - medical interventions are required for that information. The simulation can provide both. The simulation may also be able to explain how the association has developed. "That reminds me of ..." Again this will reinforce the subjective nature of the experience.

The subjective experience will be evidenced by the descriptions of feelings and diagnostic information like physiology levels and personal history. Evidence can be taken further by adding goals that are moderated by the feelings. "I'm stopping now because I'm exhausted". "I am delighted that you are pleased, can we do that again?" 

\section{Testing}
Testing will be problematic. Consciousness, as subjective experience, is defined as a first person phenomenon. It could be said that the very concept is designed to be inaccessible to scientific explanation, to preserve at least one aspect of the mind that science cannot explain away. Indeed Chalmers explicitly extracts other aspects or meanings for the term consciousness, he labels them as the easy problems and leaves the mysterious subjective experience as the hard problem for philosophers to debate. It is remains a hugely influential philosophical approach. However much the philosophical arguments are countered or neuroscience evidence accumulates, it is the subjective experience concept that continues to be presented as an objection to any AI progress. Of course the converse is also true: strict adherents of the first person perspective will not be able to refute claims of artificial consciousness. 

But there is some realistic hope. In practice assignations of consciousness are widely attributed to creatures, frustratingly without justification. And all authors assume their readers are conscious. They do so despite advocating a theory of consciousness with supposedly no behavioral characteristics, and also while denying the possibility of a reductionist approach. It is this illogical and contradictory aspect of the philosophical arguments that gives an opening to tests and demonstrations. Language and behavior will give acceptance and they are the only way a model can be tested.

Whatever the implementation or simulation the following linguistic constructs are essential:
\begin{enumerate}
\item \emph{I'm in this state so I am feeling xyz} Relates physiology to feelings, preferably supported by evidence of the state. E.g. \emph{I'm feeling tired because I'm unplugged and my battery is low} The battery state can be independently verified.
\item \emph{I'm doing this because I feel xyz}.  Relates the feelings to goals and behavior. E.g. \emph{I'm turning down screen brightness because I feel tired}
\item Regarding a stimulus \emph{that makes me feel xyz} relates a feeling to the stimuli that caused it, preferably supported by evidence of the resultant physiology. And possibly explained by how that association has been formed - some insight into the learning process and consequential associative memory. E.g. \emph{It's 9am and I'm at the station, I am feeling a bit tired} Frequently this time and place has led to a long period before being charged. You might say that my battery level is high but it does not feel like that. This distinguishes a physical battery level to a feeling that the battery level is inadequate. An extreme example is a person with a high fever being hot, but feeling cold and shivering. 
\end{enumerate}

The more that feelings are related to goals and consequential behavior, the more feelings will appear to be real. And the existence of feelings and subjective experience becomes not just a plausible explanation, but also the simplest one. 

\section{Conclusion}
We have taken the hard philosophical model of consciousness, called subjective experience. We have added to the neurological explanations and extracted a generic model. There are many claims and arguments that this is not possible. In the appendix we explain how this model is a counterexample to the main philosophical arguments. We have also added the extra features of language, goals and behavior so that the models can be readily demonstrated and tested.

Human consciousness can be simulated. Artificial consciousnesses can be simulated and explained. With specific hardware a conscious computer could be created and demonstrated. It could be argued that fiction, films, even cartoons, show that the public will recognise language and behavior and label it as consciousness.

The cultural and academic relevance is profound. But as a utilitarian feature to enhance computer systems it is hard to see any value.
We can look at evolution and consider feelings as an essential component of a primitive goal seeking control system. A system that works without abstract thought or logical analysis. We might owe our human existence to this structure, but the power of logical analysis surely makes it redundant in artificial systems.

\appendix{Counterexamples}
We have a model of human consciousness that can be simulated  and implemented to some degree. There are claims that this is impossible and such a model cannot be found. Whether the reader regards subjective experience as the essence of consciousness or not, it is interesting and instructive to relate the model to prevailing arguments that it is not possible to construct a scientific model of consciousness. The strength of the intractable philosophical problem of consciousness is not just the puzzle of what it feels like, but various well-known arguments why it cannot be explained by science. The following arguments are certainly controversial, and they have been criticized and challenged. Despite criticisms and biomedical research these arguments persist to this day. Given that we have a specific model we can either support them or offer a counterexample.
\subsection{Penrose}

\citet{penrose1991} argues that we cannot explain consciousness with information processing algorithms running on computers. Regardless of how convincing his arguments are, we agree. We reach the same conclusion. The consciousness of subjective experience requires an intimate link between the associative memory in our brain and the physiology of our bodies. Storing recalling and processing information is not sufficient. The model links our physical selves with our mental states. 

\subsection{Chalmers' twin}
\citet{chalmers1997} constructs an argument based on the logical possibility (actually, he makes an argument that is lengthy, detailed and contains many subtleties) that he has a twin, physically identical but without consciousness. The conceivability of such a twin is considered as proof that a physical explanation cannot therefore distinguish the conscious from the unconscious individual. We claim that this argument is intrinsically a circular argument, it admits simple counterexamples and it is arguably inconsistent with itself.

We agree with \citet{minsky} that Chalmers twin argument fails because it is a circular argument. The logical possibility of Chalmers' twin, physically identical but not being conscious, presumes consciousness does not have a physical explanation. If a physical explanation for consciousness existed, then either the twin could not exist or would necessarily have different physical characteristics. We offer a physical explanation. Associative memory coupled with physiology gives rise to subjective experience. Any individual with those characteristics would be conscious and a twin with an identical history, would have the same subjective experiences.

Our model answers Chalmers simply. This is a physical model of consciousness. Your twin without consciousness is not possible.

\subsection{Nagel's feeling like a bat}
\citet{nagel1974} uses a perverse argument that if we could understand the consciousness of a bat then we would have the same feelings as a bat. Since this is clearly impossible for a human, he concludes that we cannot understand consciousness. For all other phenomena an understanding would explain why we can or cannot reproduce it. Understanding how a bird flies, does not imply that we will then be able to fly. Quite the opposite, it explains why we cannot fly (Our power to weight ratio is too low). We do not agree that subjective experience is fundamentally different as a phenomenon.

Our model applies to humans, but because it is generic in nature, it can apply to other, very different individuals. It is universal. The model applies to individuals with very different physiology and explains why we cannot feel like a bat. Our physiology is different. In that we agree with Nagel. The associative memory of a bat records and recalls very different physiological responses. Responses that would be impossible for a human.

What we have in common with Nagel is an agreement that we cannot have experiences that feel like the experiences a bat has. We cannot have the same subjective experiences as a bat. The different physiology does indeed prevent that, but the model does apply to a bat, it does explain how bats will have feelings associated with their experiences.

\subsection{Color inversion}
Color inversion is another Chalmers’ argument related to the Zombie argument but more limited in scope. It is argued that a person could have red and blue experiences inverted without there being any physical change. Hence the physical details cannot explain personal experiences. In our model a person could have red and blue experiences inverted. If their associative memory connected the visual stimulus of blue with a history of warm pleasurable events and then recalled, to some extent, the hormones and physiology of warmth and comfort. That would be the opposite of many people, but quite possible.

However, the model only achieves color inversion due to physical differences. We claim that the inversion of color experiences is not possible without corresponding physical changes in the brain or body. The alternative is absurd. That a person has a red stimulus and associative memory recalls warm comfortable experiences. The warm comfortable physiology is recalled. For example, dopamine levels increase. The person would have the physiological symptoms of warmth. Could they then feel cold? It is not just implausible, the semantic definition of feeling cold is having the physiology associated with low temperatures.

The argument only has some credibility because the physiological changes on seeing a color are relatively weak. If the argument were applied to heat or cold, or Pavlov's dogs with pleasure and fear, it would be clear that the subjective experience is not independent of the physical changes in the physiology.

\subsection{Mary the monochrome neurologist}
Another argument against a physical explanation of subjective experience is the concept of Mary the neurologist who has a full understanding of sensory systems but has herself lived in a monochrome world. Even though she knows everything about the neurology associated with seeing red she has never experienced it. Does seeing red for the first time give Mary new knowledge? If so, it is argued that consciousness cannot be fully explained from physical facts.

We would draw a distinction between Mary who knows and understands the theory of seeing colours, vs. a Mary who can also make perfect real-world predictions based on that understanding. In science the distinction is quite normal. The world expert of projectile flight might know and understand all the equations for a flight of a ball or dart, but if he throws a ball, he will not know exactly where it will land. He will not know if his dart will hit the target. The real world is complicated, and we rarely know the initial conditions precisely enough to make an exact prediction, but that is not a fault or shortcoming of the theory. The same applies to Mary.

Our interpretation is as follows. Mary can have a full understanding of seeing color and the experience of seeing colour. Her understanding is based on our paper (which she much admires) augmented with further detail about the mechanisms and application in humans. She sees red for the first time and has some physiological changes which make her feel a certain way. She could not necessarily predict the exact experience, but she appreciates that it is compatible with the theory and explained by it.

Another Mary who can make perfect real-world predictions is unrealistic. It is, for all practical purposes, impossible. She would need to know the exact pattern of associative memory in her brain, all the pathways of the nerves and their connecting synapses. And also how, and to what degree, they affect her physiology. Then she would need to know the pattern of associative memory that responds to the physiology to help her say what it feels like. She might even be able to simulate it exactly by precisely inducing the predicted signals at some stage in the process. This is absurdly unrealistic in practice. Even more so when you consider that the knowledge is itself part of the associative memory. This Mary could know in advance what it would feel like to see red for the first time. If we could control our neurological and physiological signals as precisely as NASA controls the projectile flight of its spacecraft, then it may be possible to simulate new feelings before they are experienced directly.
%


\end{document}